\documentclass[usenatbib,usegraphicx]{mn2e}
\bibpunct{(}{)}{;}{a}{}{,}

\def\gs{\mathrel{\raise0.35ex\hbox{$\scriptstyle >$}\kern-0.6em 
\lower0.40ex\hbox{{$\scriptstyle \sim$}}}}
\def\ls{\mathrel{\raise0.35ex\hbox{$\scriptstyle <$}\kern-0.6em 
\lower0.40ex\hbox{{$\scriptstyle \sim$}}}}
\defcitealias{couchw87}{CS87}
\defcitealias{zabludoffa96}{Z96}

\title[Spectroscopy of E+A galaxies in AC~114]
{Spatially Resolved Spectroscopy of the E+A Galaxies in the $z=0.32$ Cluster AC~114}
\author[Michael.~ B.~ Pracy et al.]{
\parbox[t]{\textwidth}{
       Michael B.~Pracy$^1$, Warrick J.~Couch$^1$, Chris Blake$^{1,2}$, Kenji Bekki$^1$, 
 Craig Harrison$^3$, Matthew Colless$^4$, Harald Kuntschner$^5$ and Roberto de Propris$^6$}  
\\
\vspace*{6pt}\\
$^1$School of Physics, University of New South Wales, Sydney NSW 2052, 
Australia \\
$^2$ Department of Physics \& Astronomy, University of British Columbia, 
6224 Agricultural road, Vancouver, B.C., V6T 1Z1, Canada \\
$^3$Mount Stromlo Observatory, The Australian National University, Weston Creek, ACT 2611, Australia \\
$^4$Anglo-Australian Observatory, PO Box 296, Epping, NSW 2111, Australia \\
$^5$Space Telescope European Coordinating Facility, European Southern Observatory, 
Karl-Schwarzschild-Str 2, 85748, Germany \\
$^6$Astrophysics Group, HH Wills Physics Laboratory, University of Bristol, Tyndall Avenue, BS8 1TL, UK
}

\date{Received 0000; Accepted 0000}

\pagerange{\pageref{firstpage}--\pageref{lastpage}}
\pubyear{2004}
\begin{document}

\maketitle

\label{firstpage}
             
\begin{abstract}
We present spatially resolved intermediate resolution spectroscopy of a sample of
twelve E+A galaxies in the $z=0.32$ rich galaxy cluster AC~114, obtained with the
FLAMES multi-integral field unit system on the European Southern Observatory's VLT.
Previous integrated spectroscopy of all these galaxies by \citet{couchw87} 
had shown them to have strong Balmer line absorption and an absence of 
[OII]$\lambda$3727 emission -- the defining characteristics of the``E+A'' spectral
signature, indicative of an abrupt halt to a recent episode of quite vigorous
star formation. We have used our spectral data to determine the radial
variation in the strength of H$\delta$ absorption in these galaxies and hence 
map out the distribution of this recently formed stellar population. Such information
provides important clues as to what physical event might have been responsible for 
this quite dramatic change in star formation activity in these galaxies' recent past.
We find a diversity of behaviour amongst these galaxies in terms of the radial
variation in H$\delta$ absorption: Four galaxies show little H$\delta$
absorption across their entire extent; it would appear they were misidentified as
E+A galaxies in the earlier integrated spectroscopic studies. The remainder show
strong H$\delta$ absorption, with a gradient that is either {\it negative}
(H$\delta$ equivalent width decreasing with radius), {\it flat}, or {\it positive}. 
By comparing with numerical simulations we suggest that the first of these 
different types of radial behaviour provides evidence 
for a merger/interaction origin, whereas the latter two types of behaviour are more
consistent with the truncation of star formation in normal disk galaxies with the 
H$\delta$ gradient becoming increasingly positive with time after truncation.  It would seem
therefore that more than one physical mechanism is responsible for E+A formation in the 
same environment.   

\end{abstract}


\section{Introduction}
``E+A'' galaxies exhibit strong Balmer absorption lines superimposed upon an elliptical 
galaxy type spectrum \citep{dresslera83} and represent a population of galaxies that have 
undergone a significant change in their star formation rate. The lack of optical 
emission lines, for example [OII]$\lambda$3727, indicates that star formation has
ceased. However, the strong Balmer absorption line signature implies the existence 
of a substantial population of young A stars which must have formed no more than
$\sim$1\,Gyr ago. This spectral signature is usually interpreted as the recent truncation 
of a starburst 
(\citealt{dresslera83}; \citealt{couchw87}, hereafter \citetalias{couchw87}; \citealt{bargera96}; \citealt{poggiantib99}) 
or the abrupt truncation of more normal star formation activity in a disk 
galaxy \citep[\citetalias{couchw87};][]{baloghm99}.   
\par
The numbers of E+A galaxies seen in different environments evolves strongly
with redshift.  E+A's are rare in nearby clusters, making up only about 1\% of the total 
galaxy population in the clusters \citep{fabricantd91}, and are even less common
in the low-redshift field, comprising roughly 0.03\%-0.2\% of the overall galaxy 
population depending on the selection criteria
(\citealt{zabludoffa96}, hereafter \citetalias{zabludoffa96}; \citealt{blakec04}).
In contrast, these galaxies represent a significant fraction of the galaxy population in 
intermediate redshift clusters, where they were initially discovered by 
\citet{dresslera83}. Estimates of the fraction of E+A galaxies in such  
(0.2$<$z$<$0.6) clusters generally range from 10\%-20\% 
\citep{dresslera92,couchw98,dresslera99}, although
this high fraction may not be a universal property of all clusters at these redshifts
\citep{baloghm99}.
\par
The physical processes involved in the formation of these galaxies remain unclear; 
however, it is unlikely that such a dramatic change in star formation activity could be
due to internal factors and hence external `environmental' influences would seem to be
the cause. In this regard, a plethora of possible mechanisms have been suggested 
including major mergers \citep{Mihosj96},
minor mergers and galaxy-galaxy interactions
or, in the case of E+A galaxies residing in clusters, interaction with the strongly 
varying cluster tidal field \citep{bekkik99}, 
galaxy harassment \citep{mooreb98}
or interaction with the hot intracluster medium \citep{gunnj72}. There is mounting 
evidence that E+A galaxies in the low redshift field are the result of 
mergers or tidal interactions.  \citetalias{zabludoffa96} inspected ground based 
imaging of a sample of 21 E+A galaxies drawn from the 
Las Campanas Redshift Survey (LCRS), and found a high incidence of tidal features 
in these galaxies implying mergers or galaxy-galaxy interactions
in the formation process. This conclusion was later verified using high resolution HST imaging \citep{yangy04}.  
\citet{blakec04} arrived at a similar conclusion using a larger sample drawn from 
the Two Degree Field Galaxy Redshift Survey \citep[2dFGRS; ][]{collessm01}. \citet{gotot05}
using a sample of 266 E+A galaxies from the Sloan Digital Sky Survey \citep{abazajiank04} claimed to observe an 
excess in the projected local galaxy density on scales of $<100$kpc surrounding the E+A galaxies - indicating 
dynamical interactions as a likely formation mechanism. 
\par
\citet{nortons01} obtained
long slit spectroscopy of the \citetalias{zabludoffa96} sample 
in order to probe the spatial and kinematic distribution of the 
stellar populations. In most cases they found the young stellar population 
to be pressure supported, consistent with a merger
origin. The study of 2dFGRS E+A galaxies by
\citet{blakec04} supported this picture by demonstrating that these
galaxies are preferentially spheroidal systems, as evidenced by their
morphologies, group environments and luminosity function.

In contrast to the field E+A galaxies the picture for E+A galaxies 
that reside in the dense cluster environment is less clear.  The HST studies
of \citet{couchw94,couchw98} and \citet{dresslera99} revealed that the 
majority of E+As in intermediate redshift clusters are undisturbed
elliptical or early type disk systems. The very existence of such disk systems 
rules out equal mass mergers that would disrupt the disk \citep{dresslera99}, although 
minor mergers with a gas rich dwarf remains a plausible formation mechanism \citep{bekkik01}. 
In the one example of spatially resolved spectroscopy of a
high redshift ($z\approx 0.18$) cluster E+A, \citet{franxm93} found evidence for 
strong rotation in the young stellar population. 
\par
Knowledge of how this population of recently formed stars in E+A galaxies is distributed
{\it spatially} is critical information  
for differentiating between the proposed formation mechanisms. Mergers and tidal interactions lead to 
a centrally concentrated burst of star formation
\citep{Mihosj96,bekkik05} with the starburst contained within 
the central 1-2 kiloparsecs.  In contrast, ram pressure stripping should
lead to a roughly uniform truncation of star formation across the disk \citep{rosej01}.  
Using a combination of long slit spectroscopy and broadband imaging \citet{rosej01} studied the spatial distribution
of recent star formation in early type galaxies in three low redshift clusters and 
found that the star formation was centralised in comparison to the distributions observed
in similar galaxies in the field.
\citet{bartholomewl01} measured the colour gradients of early-type 
galaxies in a $z\approx 0.33$ cluster and found that the E+A galaxies tended to 
have bluer nuclei than the overall galaxy population, implying systematically more centralised star formation.      
\par
In this paper we present new data on the spatial distribution of H$\delta$ absorption  
-- one of the key tracers of the young stellar population -- for a sizable sample
of E+A galaxies in a rich intermediate redshift cluster. In contrast to previous studies, 
ours is the first to obtain such information via integrated field unit (IFU) 
spectroscopy on an 8m telescope (the VLT). If mergers and tidal interactions are 
responsible for the E+A galaxy signature, then a strong {\it negative} radial gradient 
in H$\delta$ equivalent width is expected, in the sense that the H$\delta$ equivalent width 
decreases with galacto-centric radius \citep{bekkik05}. However, if the E+A signature is the product 
of the truncation of normal star formation, a more uniform distribution of H$\delta$ equivalent width across 
the galaxy is expected. Our target sample comprises 12 E+A galaxies that are members
of the $z=0.32$ rich cluster AC~114 \citep[also known as Abell  S1077;][]{abellg89}, which 
has been the subject of many previous studies as a result of its significant
population of blue `Butcher-Oemler' galaxies \citepalias{couchw87}. 
It has a high velocity dispersion $\sigma=1660{\rm km s^{-1}}$ \citep{mahdavia01} and
an irregular X-ray morphology \citep{defilippise04}.
\par
The layout of this paper is as follows: In Section 2 we provide all the details relevant 
to the observational data used in our study, describing how our E+A sample was
selected, giving details of our IFU spectroscopic observations, outlining how the
data were reduced, and discussing difficulties encountered with the position of the
IFU's on our target galaxies. 
In Section 3 we first examine our spectra in their integrated form (i.e., summed over
the entire galaxy), comparing them with the original integrated spectroscopy obtained
by \citetalias{couchw87} and thus verifying these galaxies' classification as E+A types
based on this previous lower resolution and signal-to-noise spectral data. 
In section 4 we then exploit the spatially resolved nature of our spectroscopy to map
out the location of the young stars within our E+A galaxies, using the equivalent
width of the H$\delta$ absorption line, as defined by the H$\delta_{\mathrm{F}}$ 
index \citep{wortheyg97}, as our tracer. In particular, we determine how
H$\delta_{\mathrm{F}}$  varies with radius across each galaxy, taking into account the convolving
effects of astronomical seeing. We discuss and interpret our results in section 5, 
using model predictions for the radial gradients in metallicity and H$\delta$ strength
under different formation scenarios as our main tool. Our main conclusions are summarised
in section 6. Throughout this paper we adopt an 
$\Omega_{M}=0.3,\Omega_{\Lambda}=0.7$ and $H_{0}=70 \;{\rm kms^{-1} Mpc^{-1}}$ cosmology, which places AC~114 at 
a comoving distance of 1231\,Mpc, with $1\arcsec$ projecting to 4.6\,kpc.
\section{Data}

\subsection{Sample Selection}
We targeted the members of AC~114 which had been previously identified as E+A galaxies
in the spectroscopic study of this cluster conducted by \citetalias{couchw87}.
We chose the brightest ($R\leq 20.3$) galaxies from the \citetalias{couchw87} 
sample, and paid no attention to their colour in our selection. This meant we 
included both the reddest E+A galaxies -- which \citetalias{couchw87} referred to as ``red 
H$\delta$-strong'' (HDS) galaxies and have enhanced Balmer line absorption even though
their colours are as red as passive E/S0 galaxies -- as well as those with bluer colours
-- which \citetalias{couchw87} referred to as ``post-starburst'' (PSG) galaxies, since they are presumably seen
much sooner after the cessation of star formation. All of these galaxies have no
detectable [OII]$\lambda$3727 emission. Also included in our sample was the blue
galaxy CN667, for which the \citetalias{couchw87} study was able to confirm its membership of AC~114 but
not its spectral type. This gave us a sample of 12 galaxies. 
\par
In Table 1 we list the photometric, spectroscopic, and morphological
properties of this sample: column (1) gives the galaxy's identity number from the
original \citet{couchw84} catalogue, column (2) gives 
the $R$-band magnitude, column (3) gives the $HST$-based Hubble types determined by \citet{couchw94,couchw98}, 
column (4) gives the PSG/HDS spectral sub-type assigned by \citetalias{couchw87}, 
column (5) gives the $b_{J}-r_{F}$ colours, and column (6)  
gives the rest frame H$\delta$ equivalent width values (\AA) measured by \citetalias{couchw87}. Postage 
stamp images of all the galaxies taken in the F702W passband
with $HST$, are displayed in Fig.~1. As can be seen, the galaxies generally have
an early-type (E, S0, Sa-b) morphology, the exception being CN22 which has a 
peculiar morphology and was classified by \citet{couchw98} as an advanced merger. 
\setcounter{figure}{0}
\begin{figure*}
{\includegraphics[width=10.5cm,angle=270]{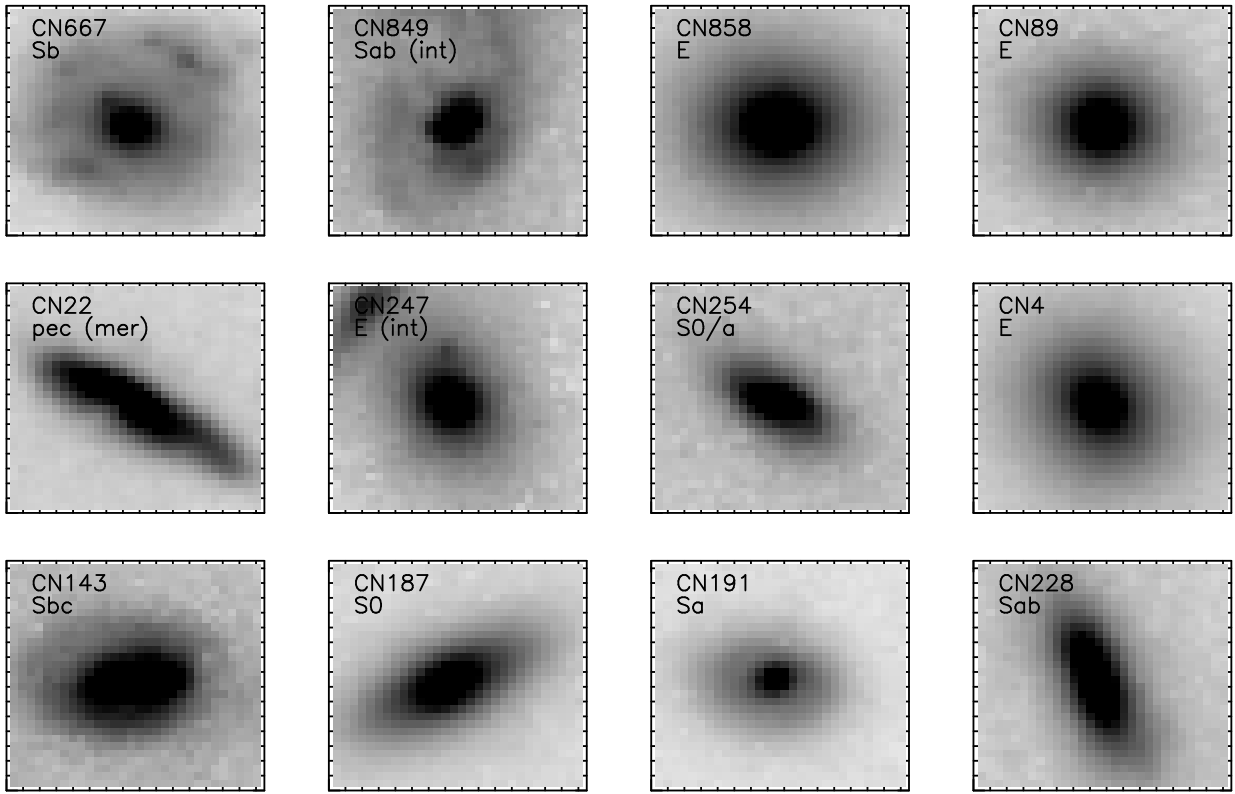}}
\caption{HST Wide Field Planetary Camera 2 (WFPC2) images of the E+A galaxy sample. Each image has dimensions $3''\times 3''$, 
corresponding to approximately 13.8 kpc at the cluster redshift and comparable to the total IFU aperture.}
\end{figure*}

\subsection{Spectroscopic Observations}
We obtained intermediate resolution IFU spectroscopy of our sample with the VLT-Kueyen telescope. 
We used the FLAMES system in the mode where it couples the OzPos multi-object 
fibre positioner to the GIRAFFE spectrograph. OzPos is able to deploy 130
single``Medusa'' fibres or 15 individual IFUs over a 25\,arcminute field at the Nasmyth
focus of the telescope. We utilised the latter to observe the twelve galaxies in our
sample simultaneously; the remaining three IFUs were used to observe blank sky regions 
in order to facilitate accurate sky subtraction. Each IFU has a $3\arcsec \times 2\arcsec$ 
($13.8 \times 9.2$ kpc) field of view which is sampled spatially by 20 square 
$0\farcs5$ lenslets; its geometry is shown in Fig.~2. 
\par
The observations were obtained during the nights of 2003 September 23, 26, and 27, 
being taken when the seeing conditions were at their best. A total of 17 exposures, mostly
of 1800\,s duration, were obtained over these 3 nights, in seeing which ranged from
$0\farcs54$ to $0\farcs84$. They gave a combined exposure time of 29,119\,s, during which
the time-weighted mean seeing was $0\farcs75$ (corresponding to a physical scale of
3.5\,kpc at the redshift of the cluster). The GIRAFFE spectrograph was used
in its `low resolution' R=6000 mode and in the ``LR4'' wavelength setting. This 
configuration allowed us to observe over the wavelength interval 
$5015{\rm \AA}\leq\lambda\leq 5831{\rm \AA}$ ($3780{\rm \AA}\leq\lambda_{\rm rest}\leq 4420{\rm \AA}$ at $z=0.32$) at
a dispersion of $\sim$0.2\AA\ per pixel.

\begin{table}
\setcounter{table}{0}
\centering
\begin{tabular}{|c|c|c|c|c|c|} \hline
CN\#& $R$ & Hubble type & Spec class & $b_{J}-r_{F}$ & H$\delta$ \citepalias{couchw87} \\ \hline
143 & 20.31 & Sbc & PSG & 1.67 & 7.7 \\
187 & 19.31 & S0 & HDS & 2.29 & 3.4  \\
191 & 19.69 & Sa & PSG & 1.61 & 5.6  \\
228 & 19.90 & Sab & PSG & 1.38 & 5.1 \\
22 &  19.73 & pec & PSG & 1.47 & 8.1 \\
247 & 19.12 & E & HDS & 2.44 & 2.8  \\
254 & 20.09 & S0/a & PSG  & 1.99  & 6.5  \\
4 & 18.44 & E & HDS & 2.29 & 4.7  \\
667 & 19.37 & Sb & ? & 1.67 & ?  \\
849 & 19.71 & Sab & PSG & 1.82 & 4.6  \\
858 & 19.12 & E & HDS & 2.35 & 3.8  \\
89 & 19.61 & E & HDS & 2.23 & 5.6  \\ \hline
\hline
\end{tabular}
\caption{Summary of the photometric, spectroscopic and morphological properties of our sample. See text for details.}
\end{table}

\subsection{Data reduction}
The data were reduced using standard IRAF routines. In brief, each frame was first
bias-subtracted and then its overscan region removed. All science frames from a given 
night were then median combined, after having been scaled to allow for the different
exposure times. Cosmic ray rejection was performed as part of this process using the 
{\sc crreject} algorithm. The flat fields were used to identify the position of the 
spectra on the CCD. A smoothed scattered light signal was removed from 
the flat field and science frames by fitting a Chebyshev function to the pixels between 
the fibres. Following this, the spectra were extracted and wavelength calibrated using
arc lamp exposures.  The flat field spectra were used to calculate and correct
for the variations in throughput from fibre to fibre; this was based on the total
signal contained within each spectrum over the entire wavelength interval 
covered by our observations. Finally, all the spectra recorded through the three
`blank sky' IFUs were summed and averaged to form a mean `sky' spectrum, which was
then used to subtract the sky from the spectra recorded in the remaining 12 `target'
IFU's. The spectra were corrected to a relative flux scale using an ARGUS-IFU observation 
of a flux standard star.

Systematic errors in the sky-subtraction will lead to incorrect continuum levels in the spectra and 
hence to biased equivalent width measurements.  In order to evaluate the accuracy of 
the sky subtraction we used the residuals of the individual `sky' 
spectra after subtraction of the `mean sky' signal.  Using the residual in the $\lambda 5577$ atmospheric emission line we calculated
the fractional error in the sky subtraction to be 3.6\%, this compares with 1.1\% expected from Poisson statistics.
More importantly, we checked for any systematic errors in the sky subtraction using the residual continuum. We calculated 
the RMS of the mean value of the residuals in each lenslet and the mean RMS of the residuals averaged over all lenslets, the 
ratio of these gives the size of any systematic errors in the sky subtraction 
with respect to the size of purely random errors. The resulting ratio
is 0.14 implying that any systematic sky subtraction error is roughly an order of magnitude smaller than the random errors.   

\subsection{IFU-Object placement}
At this point of having derived accurately sky-subtracted spectra for the 20 lenslets
in each IFU, it became apparent that most of our target galaxies were not perfectly 
centred in each IFU. An example is shown in Fig.~2 where we plot the spectra 
obtained in each of the IFU `pixels' for the galaxy CN191. Note that the normalisation
of each spectrum is identical, so the continuum level of the twenty spectra can be
directly compared. We see quite clearly in this case that the signal is at a maximum 
in the two spectra immediately below the centre of the IFU, indicating that the galaxy 
is off-centre in this direction by $\sim 0\farcs25-0\farcs50$.  
\par
In Fig.~3 we display the centroid positions of each galaxy within the IFU field, estimated 
from the total galaxy signal observed in each IFU pixel (see Section 4.2). The centroid 
positions of each galaxy in the coordinate system of Fig.~3 are listed in columns (5) and (6) 
of Table 2. It can be seen from Fig.~3 that the galaxies are generally
off-centre by at least half an IFU pixel ($0\farcs25$), with quite a number being
off-centre by as much as 1-2 IFU pixels ($0\farcs5-1\farcs0$). This could be due to 
one or more of the following factors: (i)\,errors in the astrometric positions 
used for our galaxies, (ii)\,IFU positioning errors, and (iii)\,errors in the determination 
of the galaxy centroids from our IFU data.   
\par
Since the placement of the IFU's on target objects involves bringing them in approximately
radially from their parked locations around the perimeter of the field, the IFU's
will generally all be at different position angles with respect to the sky. Hence Fig.~3
gives no sense as to whether the three possible sources of error above are random or
systematic. In Fig.~4, therefore, we plot the observed positional offsets in terms of their 
magnitude and direction on the sky. The vectors point in the direction of the calculated 
centre of the galaxy from the geometric centre of the IFU, and their lengths 
have been scaled up by a factor of 100 for display purposes. It would appear from this 
diagram that the positional errors are systematic on the sky, with perhaps marginal evidence 
that their magnitude increases with distance from the centre of the field. This implies there is 
some systematic problem with the IFU positioning across the field, either
with the OzPos positioner or, more likely, in there being a systematic positional 
offset between the astrometric zero points for our
guide stars (which are not plotted in Fig.~4) and that of our target galaxies. As
disappointing as it may be, we are left to conclude that this problem is most likely due 
to a combination of all three of the different types of error mentioned above, although
we note that repeated tests of the OzPos positioner have indicated it is capable of
placing the IFU's to a RMS precision of better than $0\farcs2$. How we deal with this
problem in our subsequent analysis is further detailed in section 4.2.  

\setcounter{figure}{1}   
\begin{figure}
{\includegraphics[width=6.0cm,angle=270]{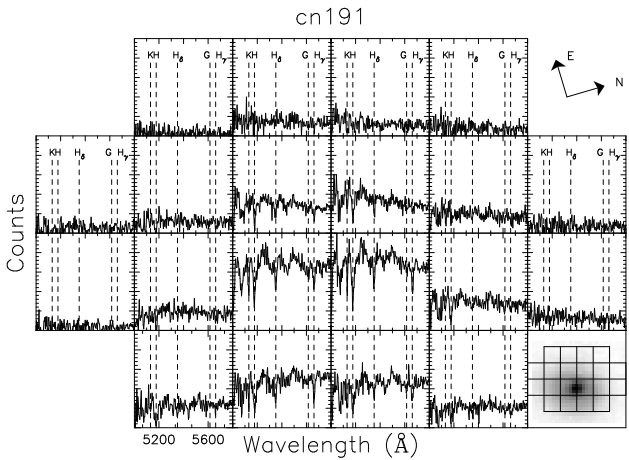}}
\caption{Spatial geometry of the IFU's used in our study. Each square represents an 
IFU lenslet, which is $0\farcs5$ on a side. The spectrum obtained through that 
lenslet (for CN191) is displayed within. The ordinate scale is identical in all 
lenslets; thus the continuum level reflects the level of light observed in that direction on the sky. 
Note the astrometric offset, which is discussed in Section 2.4.
The bottom right panel shows the $HST$ image of CN191 and its orientation with respect 
to the IFU. }
\end{figure}
\setcounter{figure}{2}
\begin{figure}
{\includegraphics[width=7.0cm,angle=270]{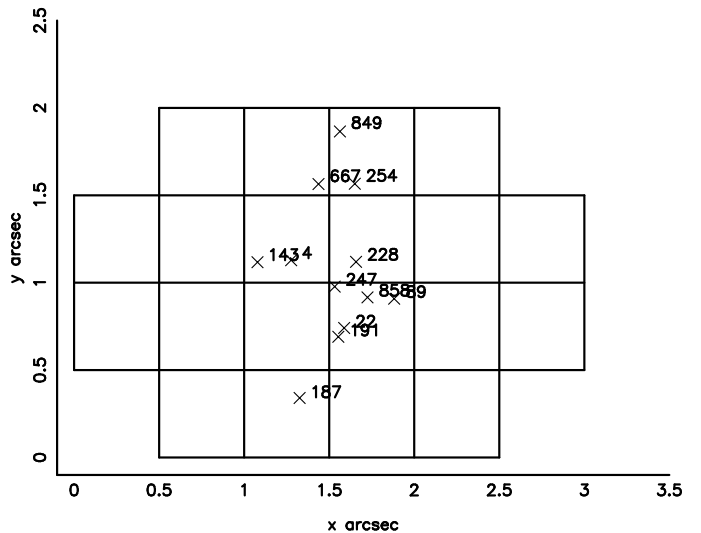}}
\caption{The centroid positions of our galaxies within the IFU field; objects are  labelled with their CN number.}
\end{figure}
\setcounter{figure}{3}
\begin{figure}
{\includegraphics[width=6.5cm,angle=270]{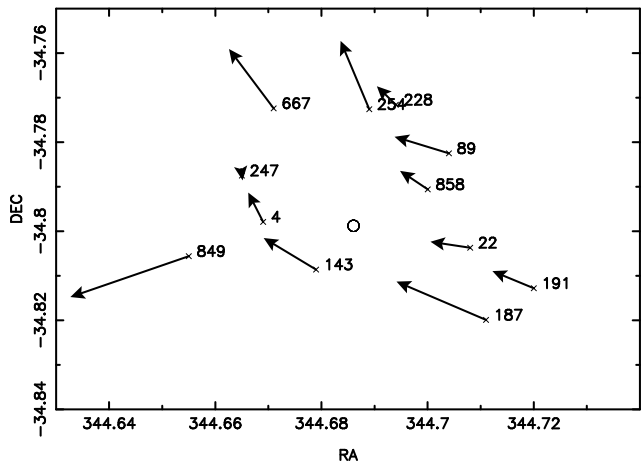}}
\caption{Vectors showing the magnitude and direction of the positional offsets of our
target galaxies within the IFU's, as they appear on the sky. The lengths of the vectors 
have been scaled by a factor of 100 for display purposes. The {\it circle} indicates the centre of the FLAMES field of view}
\end{figure}

\begin{table*}
\setcounter{table}{1}
\centering
\begin{tabular}{|c|c|c|c|c|c|c|c|c|c|c|c|c|c|c|} \hline
CN no. & z & $b_{J}-r_{F}$ & $K_{b_{J}-r_{F}}$& $x_{cen}$& $y_{cen}$&$r_{e}$&H$\delta$\citepalias{couchw87}& H$\delta_{\mathrm{F}}$  &  H$\gamma_{\mathrm{F}}$ &Slope (H$\delta_{\mathrm{F}}$) & corr slope \\
 & & & & arcsec & arcsec & kpc & \AA&  \AA & \AA &${\rm\AA\; r_{e}}^{-1}$ & ${\rm\AA\; r_{e}}^{-1}$ \\ \hline
143 & 0.3103 & 1.67 & 0.64 & 1.08 & 1.12 &3.2& 7.7  &$ 4.18\pm 0.59$  & $4.36\pm 0.48 $  &$0.80\pm 1.68 $ & 1.6 \\
187 & 0.3074 & 2.29 & 1.03 & 1.33  & 0.34 &2.6& 3.4 &$ -0.62\pm 0.56$  & $-2.39\pm 0.42$ &$-0.69\pm 0.67 $ & -1.8\\
191 & 0.3045 & 1.61 & 0.60 & 1.55 & 0.69 &1.9& 5.6 & $ 3.20\pm 0.38$  & $3.90\pm 0.31$  &$-2.37\pm 0.44  $ & -6.4 \\
228 & 0.3161 & 1.38 & 0.46 & 1.66 & 1.12 &3.4& 5.1 & $ 5.82\pm 0.66$  & $5.67\pm 0.56$  &$1.96\pm 0.85  $ & 5.8 \\
22 & 0.3356 & 1.47 & 0.53 &  1.59 & 0.74 &3.4& 8.1 & $ 4.50\pm 0.38$  & NA              &$-2.47\pm 0.80  $ & -5.8 \\
247 & 0.3191 & 2.44 & 1.15 & 1.53 & 0.98 &1.9& 2.8 & $ 1.65\pm 0.56$  & $-0.41\pm 0.41 $ &$-0.14\pm 0.55 $ & -0.2 \\
254 & 0.3192 & 1.99 & 0.86 & 1.65 & 1.57 &2.7& 6.5 & $ 2.24\pm 1.22$  & $2.10\pm 1.00$  &$ -1.48\pm 1.45 $ & -4.3 \\
4 &  0.3083 &  2.29 & 1.04 & 1.28  & 1.13 &5.2& 4.7 &$ -1.03\pm 0.38$  & $-1.13\pm 0.28$ &$1.30\pm 1.60 $ &  2.6 \\
667 & 0.3122 & 1.67 & 0.64 & 1.44 & 1.56 &4.4& ? &   $ 0.11\pm 0.50$  & $0.99\pm 0.38$  &$0.77\pm 1.32  $  & 1.3  \\
849 & 0.3235 & 1.82 & 0.75 & 1.56 & 1.87 &2.6& 4.6 & $ 3.64\pm 0.30$  & $0.75\pm 0.24$  &$0.84\pm 0.40  $ & 2.1 \\
858 & 0.3118 & 2.35 & 1.00 & 1.73 & 0.92 &3.8& 3.8 & $ -0.77\pm 1.15$  & $-1.51\pm 0.88$ &$2.17\pm 2.03 $ & 6.1 \\
89 & 0.3173 & 2.23 & 1.01 & 1.88 & 0.91 &2.8& 5.6 &  $ 2.60\pm 0.65$  & $-0.77\pm 0.57$ &$0.39\pm 0.89  $ & 0.9 \\ \hline
\hline
\end{tabular}
\caption{Columns from left to right: galaxy ID, redshift, observed $b_{j}-r_{f}$ colour, K-correction, galaxy 
centres in the x-y reference frame defined in
Fig.~3, effective radius measured from the HST images, H$\delta$ equivalent width from \citetalias{couchw87}, H$\delta_{\mathrm{F}}$  from spatially 
integrated FLAMES spectra, H$\gamma_{\mathrm{F}}$ from spatially integrated FLAMES spectra , slope of the H$\delta_{\mathrm{F}}$  radial profiles (normalized to the effective radius) and the radial profile slopes corrected for the effects of astronomical seeing. Note the redshift of CN22 precludes the measurement of the H$\gamma_{\mathrm{F}}$ index.}
\end{table*}

\section{Spatially integrated spectra}

\subsection{Construction and measurement}
Before utilising the spatially resolved spectroscopic information provided by our 
observations, it is instructive to first examine it in its `integrated' form and
see how our spectra compare with those previously obtained by \citetalias{couchw87}. For each galaxy, 
we combined the 20 individual spectra obtained with each IFU into a single 
spatially integrated spectrum. Each spectrum was weighted according to its variance 
to maximise the signal-to-noise ratio of the final integrated spectrum. The spectra for the 
12 galaxies are presented in Fig.~5, having been smoothed with a Gaussian of FWHM  
1.2\AA. It can be seen that these integrated spectra are of reasonably high quality with ${S \over N}$
ranging from $\sim$5[\AA$^{-1}$] for CN254 to $\sim$20[\AA$^{-1}$] for CN849, measured in a 100-\AA-wide interval
(rest frame) just redwards of the H$\delta$ feature.  
Their signal-to-noise ratios are certainly higher than those of the original spectra
obtained by \citetalias{couchw87}. As such, the key spectral lines within our observed wavelength
range (indicated by the vertical {\it dashed lines} in Fig.~5) are easily identifiable, 
in particular the Balmer absorption lines H$\delta$ and H$\gamma$. However, these two 
Balmer lines are conspicuous by their weakness or even absence in several of the galaxies 
(CN858, 667, 4, 187), bringing into question their classification as E+A types.    
\par
These visual impressions were properly quantified by conducting redshift and spectral
line index measurements. Redshifts were measured for each spectrum using the {\sc IRAF} 
task {\sc FXCOR}, which performs a Fourier cross-correlation \citep{tonryj79} of the object 
spectrum with a set of template spectra taken from \citet{vazdekisa99}. The resulting redshifts 
for our 12 target galaxies are given in column (2) of Table 2. These redshifts are in good agreement with those originally
derived by \citetalias{couchw87} with the RMS difference between the two sets of redshifts being $\Delta z_{RMS} \approx 0.001$.
\setcounter{figure}{4}
\begin{figure*}
{\includegraphics[width=15.5cm,angle=270]{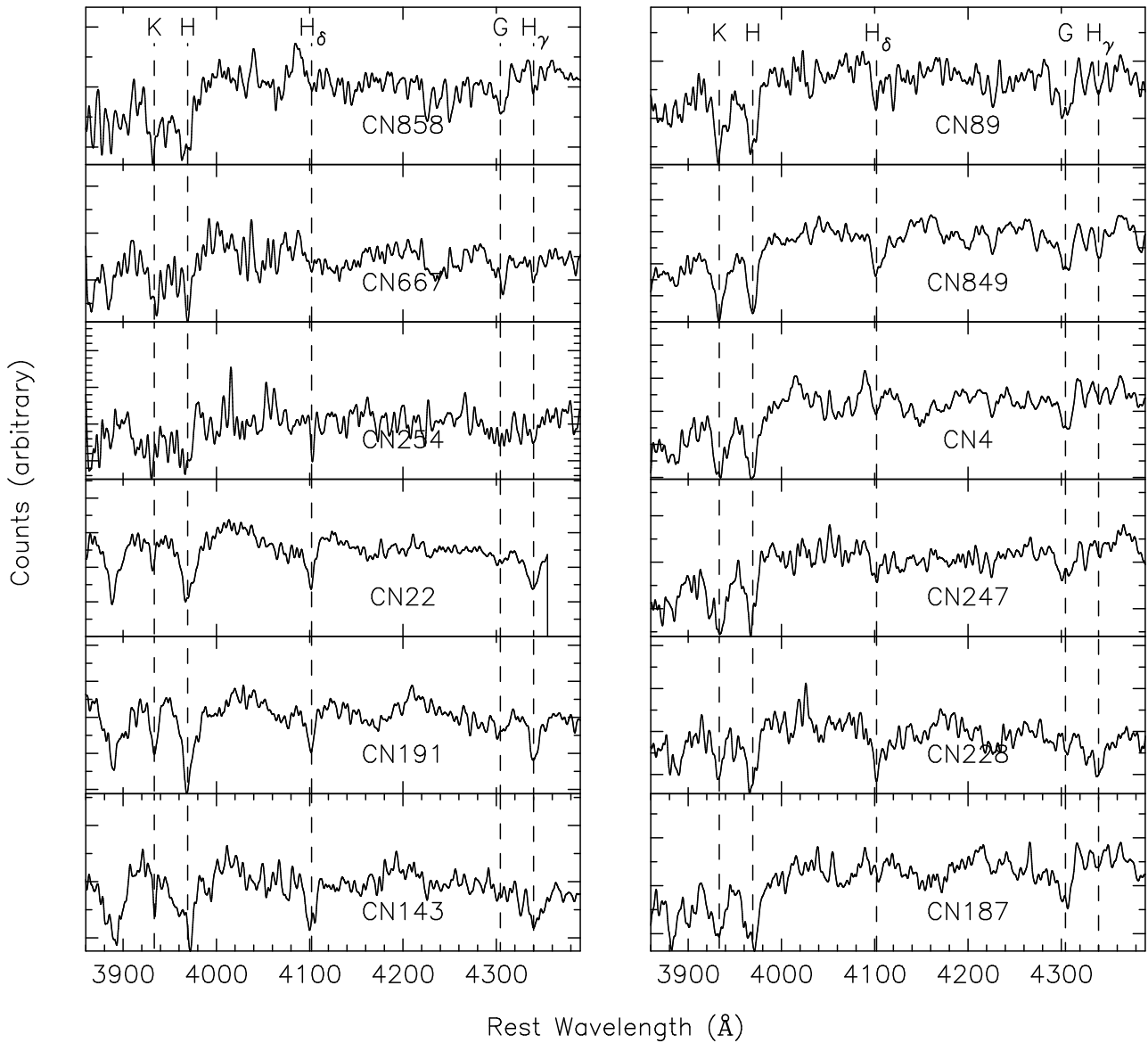}}
\caption{Spatially binned redshift-corrected spectra for the twelve galaxies in our E+A galaxy sample.}
\end{figure*}

These redshifts were then used to precisely locate spectral lines within each spectrum
and measure their equivalent width. Of primary importance here is the H$\delta$
line, given its use in defining an E+A galaxy \citep[\citetalias{couchw87};][]{dresslera99}. To
measure the equivalent width of this line we adopted the H$\delta_{\mathrm{F}}$ index 
definition of \citet{wortheyg97}, whereby the signal within the line is evaluated over 
the rest-wavelength interval 4091.00-4112.25\,\AA, and the neighbouring continuum levels are 
determined within the intervals 4057.25-4088.50\,\AA\ and 4114.75-4137.25\,\AA. The error 
in this index was calculated directly from the spectrum's variance array \citep[see][]{cardieln98}. 
The H$\delta_{\mathrm{F}}$  values derived in this way are plotted in Fig.~6 ({\it filled circles}) 
and listed in column (9) of Table~2, for comparison the H$\gamma_{\mathrm{F}}$ index values are given in column (10) 
of Table~2 and generally scale well with H$\delta_{\mathrm{F}}$.

\setcounter{figure}{5}
\begin{figure}
{\includegraphics[width=7.0cm,angle=270]{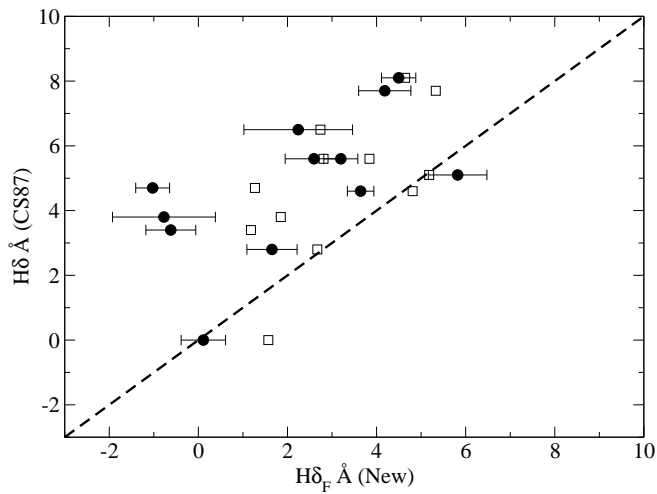}}
\caption{The H$\delta$ equivalent width values from \citetalias{couchw87} versus those measured 
from our spatially integrated spectra (shown in Fig.~5). The {\it filled circles} show
the comparison with our H$\delta_{\mathrm{F}}$ index values; the {\it open squares} show the comparison
with the values we obtain using the same interactive equivalent width measuring 
technique as that used by \citetalias{couchw87}. The point with a \citetalias{couchw87} $H\delta$ value 
of zero corresponds to CN667 for which no equivalent width measurement was made.}
\end{figure}

\subsection{Comparison with CS87}
It is worthwhile comparing our line index measurements with those of
\citetalias{couchw87}, and re-examining their E+A status. In Fig.~6 we plot \citetalias{couchw87}'s H$\delta$ 
equivalent width values against our own.
The first comparison we make is with our H$\delta_{\mathrm{F}}$ index values; this is
shown by the {\it filled circles}. A low value of H$\delta_{\mathrm{F}}$ (H$\delta_{\mathrm{F}}\approx 0$) is detected
in galaxies CN858, CN667, CN4 and CN187, confirming the visual impressions mentioned
above. As noted earlier, CN667's spectral classification was 
previously uncertain \citepalias{couchw87}, and we can now confirm 
it to be ``passive'' or, in the nomenclature of \citet{dresslera99},
a ``k'' type. Note since no \citetalias{couchw87} H$\delta$ 
equivalent width index is available for CN667, it is arbitrarily assigned an ordinate value of 
zero in Fig.~6.   The other three galaxies were, however, all classified as E+A's;
our better quality and higher spectral resolution data clearly show these classifications
to be incorrect.  
It can be seen, in Fig.~6, that our measured H$\delta_{\mathrm{F}}$ indices are generally 
smaller than the \citetalias{couchw87} values, the average difference being $\sim$2.7\,\AA. 
However, we caution that \citetalias{couchw87} used a different method for measuring the equivalent
widths of lines in their spectra, and their integrated spectra were measured through
a $2\arcsec$ aperture fibre, the exact positioning of which on each galaxy cannot be
determined.
\par
While we are unable to determine what, if any, aperture effect the latter
might introduce into our comparison, we can at least address the issue of different
line measuring techniques. To do so, we re-measured the H$\delta$ equivalent width values of our 
spatially integrated spectra using the same `interactive' routine that \citetalias{couchw87} used
(the routine {\sc ABLINE} in the Starlink software package). Here, the line and
neighbouring continuum regions are identified interactively, an iterative polynomial
fit is performed to determine the continuum level across the line, and the line
itself is fitted by a Gaussian function within the wavelength interval identified.
Since spectral resolution can affect the measured equivalent width values, we were careful to 
first smooth our spectra to the resolution of the original \citetalias{couchw87} spectroscopy.
The comparison with  the H$\delta$ equivalent width values measured using this technique 
is shown in Fig.~6 by the  {\it open squares}. It can be seen that while our
measurements are still systematically smaller than those of \citetalias{couchw87}, the difference is
nowhere near as great (cf. the H$\delta_{\mathrm{F}}$ measurements), the mean difference being $\sim$1.7\,\AA. 
Nonetheless, our comparison points to the fact that systematic
differences in line index measurements can exist between different studies and
data sets, and some caution needs to be exercised accordingly. One obvious cause 
of such differences, particularly for multi-object fibre-based studies, is a 
systematic error in sky-subtraction. This would lead to incorrect continuum levels
within the spectra, and hence an over- or under-estimation of the equivalent widths.

Hereafter, we only consider the eight galaxies in our sample where we have detected 
significant (H$\delta_{\mathrm{F}} > 1$\AA) Balmer absorption.

\subsection{Spectrophotometric models revisited}
With there having been some revisions to the measured strengths of the key Balmer
absorption lines within our E+A sample, a reexamination of their evolutionary
status is warranted. This is best done using the H$\delta_{\mathrm{F}}$--colour diagnostic
diagram \citepalias{couchw87}, where we compare the data for our galaxies with spectrophotometric
models. This is shown in the top panel of Fig.~7, where we plot the H$\delta_{\mathrm{F}}$ values measured for our 
galaxies against their $b_{J}-r_{f}$ colours (the reddening-corrected values taken from \citetalias{couchw87}).
The spectra were smoothed to a resolution of $3{\rm \AA}$ prior to measuring the H$\delta_{\mathrm{F}}$ indices
in order to facilitate a consistent comparison with model spectra. 
We have K-corrected the colours using the formulae given by Wild et al. (2004), 
which calculate a K-correction based on redshift and colour:
\begin{equation}
K_{b}=[-1.63+4.53x]y+[-4.03-2.01x]y^2-{z \over 1+(10z)^4}
\end{equation}
\begin{equation}
K_{r}=[-0.08+1.45x]y+[-2.88-0.48x]y^2 
\end{equation}
where $x=b_{J}-r_{f}$ and $y=z/(1+z)$. The magnitude of the K-correction for each galaxy is given 
in column (4) of Table 2. 
\par
Also plotted in the top panel of Fig.~7 are the evolutionary tracks for galaxies with different star
formation histories; these were calculated using the stellar population synthesis models 
of \citet{bruzualg03}. The Bruzual \& Charlot code returns line index measurements 
and $B-R$ colours at a series of time steps for the chosen star formation history computed directly 
from synthetic model spectra at $3{\rm \AA}$ resolution.  
It is important to note that the line index measurements here were made using the
same H$\delta_{\mathrm{F}}$ definition (at the same spectral resolution) that was applied 
to our observed spectra, thereby ensuring complete
consistency in comparing our observations with the models. As far as the model
colours were concerned, they were transformed into  
$b_{J}-r_{f}$ values using the conversion of \citet{couchw81}:
\begin{equation}
b_{J} - r_{f} = -0.017 + 1.059 \, (B-R) - 0.027 \, (B-R)^2
\end{equation}
In all cases we assumed a Salpeter \citep{salpetere55} initial mass function 
and an exponentially decaying star formation rate. The {\it solid curves} in the top panel of Fig.~7 trace the evolution in the 
H$\delta$-colour plane for solar metallicity models in which this exponentially-decaying star
formation is abruptly (instantaneously) truncated after 10\,Gyrs. The model curves are shown from the time
the star formation is truncated to a time 10\,Gyrs after the end of star formation.
The models displayed have exponentially-decaying 
star formation rates with e-folding times $\tau=$5, 10 and 15\,Gyrs.  
The $\tau=$15\,Gyr model has the highest value of H$\delta_{\mathrm{F}}$ and bluest colour at the time the
star formation is truncated (since it has the highest star formation rate at this time). The  $\tau=$5 and 10\,Gyr
models have progressively lower H$\delta_{\mathrm{F}}$ index values and redder colours.
The {\it dashed curve} represents a $\tau$=5\,Gyr model in which a $\delta$-function burst of star formation occurs 
after 10\,Gyrs, converting 10\% of the total galaxy mass into new stars.  The {\it dot-dashed curve} is a variation of the  
starburst model in which the starburst, converting 10\% of the galaxy mass into stars, takes place over a period of 1\,Gyr. 
In order to investigate the effect of metallicity on these evolutionary tracks in the H$\delta_{\mathrm{F}}$--colour 
plane, we also plot the tracks of the $\tau$=10\,Gyr model 
for sub-solar (Fe/H$=-0.64$) and super-solar (Fe/H$=0.56$) metallicities. These are shown as the dotted lines in the top panel of  Fig.~7. 
The lower metallicity model track is shifted to have bluer colours than in the case of solar metallicity
and the high metallicity track is shifted redward. Note the almost monotonic increase of
observed H$\delta_{\mathrm{F}}$ with increasing blueness. This is consistent with the general trends seen in
all the model tracks. However, apart from the low metallicity models, the data points in the 
top panel of Fig.~7 all lie blueward and/or beneath the model 
curves. This may be a result of errors in, any or all of, the 
reddening correction, colour conversion and K-correction
causing the data to have colours which are too blue. Emission line filling from residual star formation present in the data but
not in the models resulting in lower values of the measured H$\delta_{\mathrm{F}}$ index could also explain the discrepancy.
In the bottom panel of Fig.~7 we show the galaxies in the H$\delta_{\mathrm{F}}$--D4000 plane. We use the D4000 index definition of \citet{baloghm99} which 
measures the ratio of the flux in a band just redward of the 4000\AA\, break (4050-4150\AA) to the flux in a band just blueward of the 4000\AA\, 
break (3850-3950\AA).  This index is less affected by dust than broadband colours 
and does not rely on uncertain colour conversions and K-corrections.  The same set of models 
described above are superimposed on the data and, in this case, there is no systematic offset in the D4000 index value between 
the model tracks and observational data.   
\par
It is apparent from the tracks in Fig.~7 that it is difficult to distinguish between 
the truncation of a starburst episode and the abrupt truncation of normal star formation, unless 
the galaxy is observed shortly after the starburst when it will display bluer 
colours and stronger Balmer line absorption than can be reproduced by the truncation of normal 
star formation (upper left region of Fig.~7).{\it We are unable to differentiate between 
the truncation of exponentially-decaying star formation and the sudden cessation of a 
starburst, on the basis of the `integrated' {\rm H$\delta$} equivalent widths and broadband colours alone}. 
It is worth noting that the duration over which these
models can be distinguished is very short. For the starburst models presented in Fig.~7, 
the interval between the time star formation ceases and the model galaxy evolves
redwards of $b_{J}-r_{F}>1.0$, is approximately 600\,Myrs.  The short lifetime of this very
blue H$\delta$-strong phase implies that very few galaxies should be observed in this part 
of parameter space. 

\setcounter{figure}{6}
\begin{figure}
    \begin{minipage}{0.47\textwidth}
      \includegraphics[width=7.0cm,angle=270]{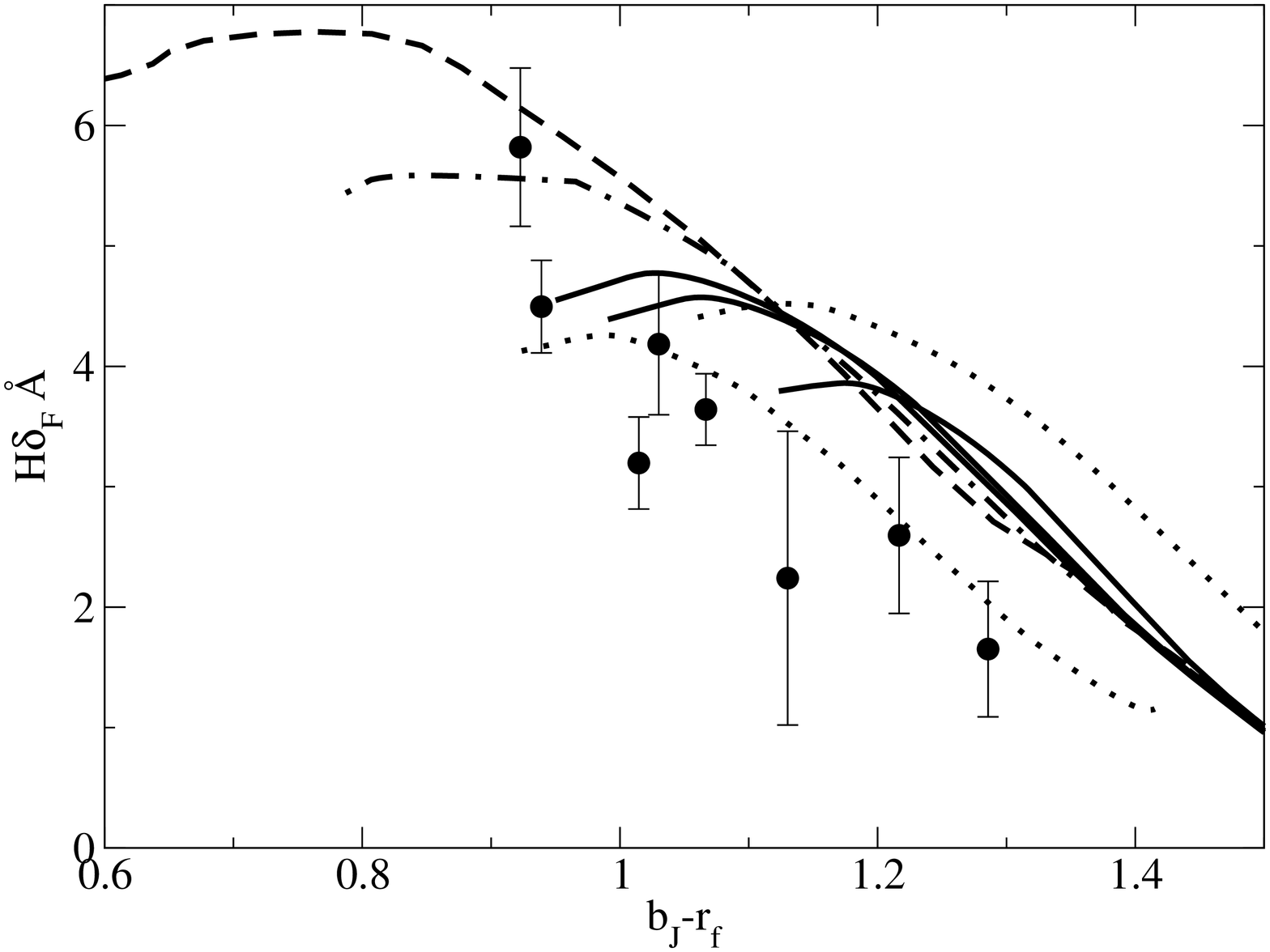}
      \includegraphics[width=7.0cm,angle=270]{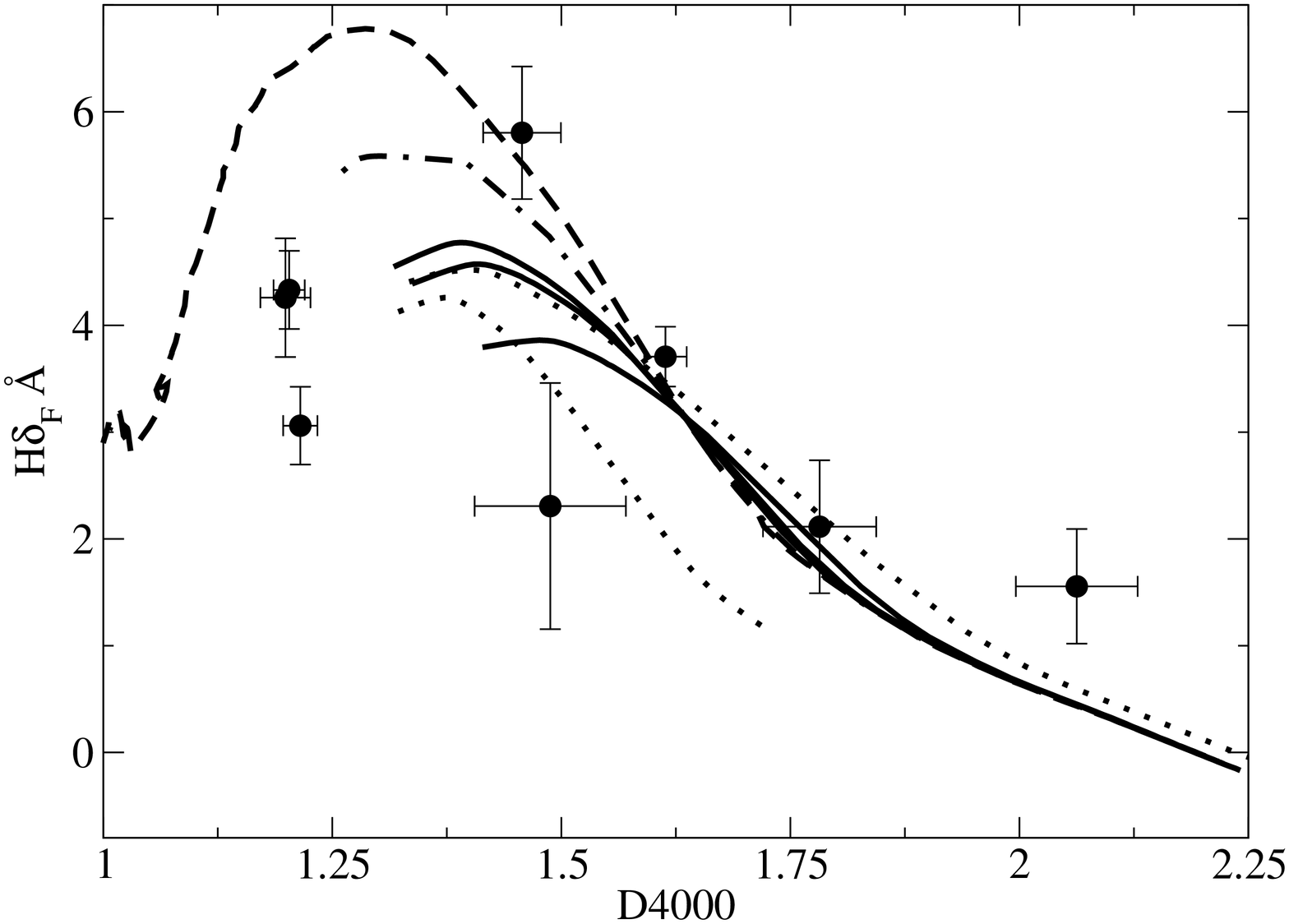}      
    \end{minipage}
\caption{Top panel: H$\delta_{\mathrm{F}}$ plotted 
against $b_{J}-r_{f}$ colour. The H$\delta_{\mathrm{F}}$ measurements from our sample are plotted as 
{\it filled circles}. The galaxy colours have been corrected for reddening \citepalias{couchw87} and K-correction. 
The curves are evolutionary tracks produced using the spectral synthesis models 
of \citet{bruzualg03}. The {\it solid lines} represent evolutionary tracks for models in 
which normal star formation is suddenly truncated. The {\it dashed and dot-dashed lines} show the 
evolution for a post-starburst galaxy. The {\it dotted lines} show the effect of varying metallicity.  
The details of these models are described in the text. Bottom panel: H$\delta_{\mathrm{F}}$ plotted 
against the D4000 index. The model curves are the same as those shown in the top panel.}
\end{figure}

\section{Spatially resolved spectra}

\subsection{Defining the galaxy centre}
Due to our target galaxies being off-centre within the IFU's (see section 2.4), our
first step prior to analysing our spatially revolved spectral data was to find  
where each galaxy was centred within the IFU. This was done by
summing the flux in each IFU element over the full wavelength range of the spectra 
to create an image in the IFU of each galaxy.  It is insufficient to simply calculate a 
flux-weighted mean position for the centre of the galaxy, because the small spatial extent of 
the IFU will cause us to underestimate the offset of the galaxy 
centre from the centre of the IFU. Instead we fitted a surface $z(x,y)$ to the IFU image: 
\begin{equation}
z(x,y)=a+bx+cx^2+dy+ey^2
\end{equation}
and required the coefficients $c$ and $e$ to both be negative.
The galaxy centre is then defined by the maximum of the function $z(x,y)$:
\begin{equation}
x_{cen}={-b \over 2c}\;\;{\rm and }\;\;y_{cen}={-d \over 2e}.
\end{equation}
Our choice of function is motivated only by the requirement that it has a well-defined maximum; the precise choice of   
functional form has little effect on the derived position of the maximum.   
In Fig.~3 we show the positions of the centre of the galaxies as determined from the maxima of the fitted paraboloids.

\subsection{Radial profiles}
The signal-to-noise ratio of the spectra obtained in each individual IFU lenslet 
is insufficient to accurately map the spectral line features at an individual IFU pixel
scale ($0\farcs5$). Hence we spatially binned the IFU spectra in annuli around the 
galaxy centres, focusing our attention on {\it radial} variations in the spectral 
properties. 

In binning the data in this way, the weight assigned to the ith lenslet was taken
to be:
\begin{equation}
w_{i}={f_{i}A_{i} \over \sum f_{i}A_{i}}
\end{equation}
where $f_{i}$ is the total flux observed through the lenslet, and $A_{i}$ is the area of 
the lenslet which falls within the annular bin (the value for which was calculated
numerically). The spectra were binned into three concentric annuli, which were 
partitioned at $0\farcs5$ and $1\farcs0$. The outermost annulus contained all 
of the IFU spectra at radii greater than $1\arcsec$, and its effective radius 
was taken as the midpoint between the $1\arcsec$ ring and the smallest ring which contained the entire 
IFU field of view. This results in a ${S \over N}$ of $\sim$2-3[\AA$^{-1}$] for the outer bins of the faintest galaxies to
a ${S \over N}$ of $\sim$16[\AA$^{-1}$] for the central bin of CN849. The H$\delta$ equivalent width measurements were then 
repeated on the binned spectra obtained in each of the three annuli. 

The radial variation in H$\delta_{\mathrm{F}}$ observed in each of our galaxies can be seen in Fig.~8, 
where we plot the values measured in the three annular bins. Radial gradients 
are seen in most galaxies, and for the purposes of our subsequent analysis 
(in particular, deconvolution; see below) we represent these via a linear fit to the data points:
\begin{equation}
H\delta_{\mathrm{F}}(r)=\alpha + \beta {r \over r_{e}},  
\end{equation}
where $r_{e}$ is the effective radius of the galaxies measured from the HST imaging and are given in column (7) of Table 2 
and the slopes, $\beta$, being listed in column (11) of Table 2. 

\setcounter{figure}{7}
\begin{figure*}
{\includegraphics[width=12cm,angle=270]{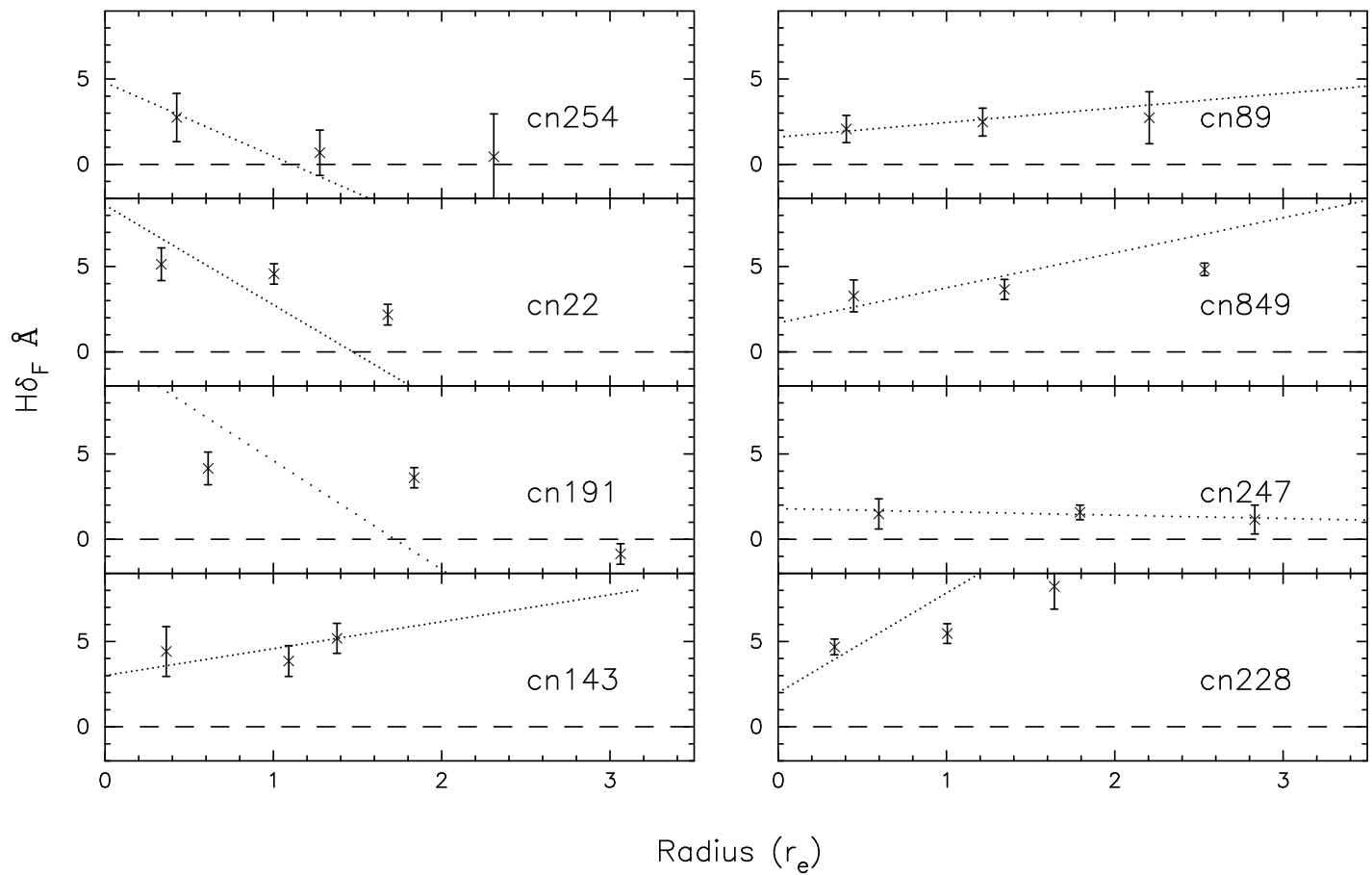}}
\caption{Radial H$\delta_{\mathrm{F}}$  profiles for the eight strong Balmer absorption line galaxies in our sample. The {\it  stars} represent the
measured H$\delta_{\mathrm{F}}$ values and the {\it dotted lines} represent the radial gradients after deconvolution with the seeing disk.
The radial distances are shown in units of the galaxies' effective radii as measured from the HST images (Fig.~1).}
\end{figure*}

\subsection{Deconvolving the seeing disk}
The galaxy flux distributions that we observe across the IFU lenslets are
convolved with a seeing disk which has a time-averaged FWHM ($0.75''$)
which exceeds the side length of a lenslet ($0.5''$). We therefore
expect the `true' deconvolved radial equivalent width profile to differ
significantly from that which is observed.  Qualitatively, any
observed equivalent width gradient will be more pronounced before convolution.  In
detail, we denote $EW_{\rm con}$ and $F_{\rm con}$ as, respectively, the equivalent 
width and flux maps after convolution with the seeing disk.
Let $EW_{\rm dec}$ and $F_{\rm dec}$ be the deconvolved maps of these
quantities.  The equations governing the convolution (which is denoted
by the overline symbol) are:
\begin{eqnarray}
F_{\rm con} &=& \overline{F_{\rm dec}} \nonumber \\ F_{\rm con} \times
EW_{\rm con} &=& \overline{F_{\rm dec} \times EW_{\rm dec}}
\end{eqnarray}
                                                                                
In light of the inherent difficulties in performing numerical
deconvolutions of general functions, we adopted the following method to
determine the `true' equivalent width profile $EW_{\rm dec}$, which we
assume is a linear function of radius $EW_{\rm dec}(r) = \alpha^{\prime} +
\beta^{\prime} \, {r \over r_{e}}$.  Firstly, we fitted an elliptical Gaussian to the pixelised
map of $F_{\rm con}$, and deconvolved this function analytically using
the observed seeing FWHM to obtain $F_{\rm dec}$.  We then looped over a grid
$(\alpha^{\prime},\beta^{\prime})$ of profiles for $EW_{\rm dec}$.  For each grid point
we deduced the resultant convolved map $EW_{\rm con} =
({\overline{{F_{\rm dec}} \times EW_{\rm dec})}/F_{\rm con}}$, which we 
binned radially for comparison with the linear fits to the 
observed [H$\delta_{\mathrm{F}}$, $r$] data points (Table 2).  The best-fitting values 
of $(\alpha^{\prime},\beta^{\prime})$ form our best model for the deconvolved radial equivalent width profile of each 
galaxy.
\par                                                                                
The revised slopes $\beta^{\prime}$ are listed for each galaxy in the final column of Table 2.  
We note that the corrections involved are substantial (a modification of the slope 
by a factor $\approx 2$), and always serve to increase the significance of the
slope detection. This can be seen visually in Fig.~8 where we have plotted
(as {\it dotted lines}) the `deconvolved' linear relations; it is important to
stress here that their behaviour is not simply determined by the radial variation
seen in the observed H$\delta_{\mathrm{F}}$ values, but also the galaxy's radial flux
distribution!

\section{Discussion}

\subsection{General trends}
Inspection of Fig.~8 (or Table 2), whether it be the observed data points or the more accentuated
`deconvolved' linear profiles, indicates three distinct types of behaviour in terms
of the variation of H$\delta_{\mathrm{F}}$ with galacto-centric radius: (i)\,A {\it negative}
slope, with H$\delta$ absorption being strongest at the centre of the galaxy and
decreasing roughly monotonically with increasing radius -- the E+A galaxies in our
sample which show this type of behaviour are CN254, CN22, and CN191. 
(ii)\,A {\it flat} slope, with H$\delta_{\mathrm{F}}$ being uniform across the entire face of the galaxy 
-- Three galaxies show this type of behaviour, and they are CN143, CN89 and CN247 (although arguably this latter 
galaxy is not an E+A type, due to its H$\delta$ absorption being too weak [H$\delta_{\mathrm{F}} < 2$\,\AA]).
(iii)\,A {\it positive} slope, with H$\delta$ absorption being at its weakest at the centre of
the galaxy and increasing roughly monotonically with increasing radius -- the galaxies
in this category are CN849 and CN228.

Curiously, however, there appears to be little correlation between H$\delta$ profile
slope and overall morphology. Two of the three negative slope galaxies are early-type
disk galaxies (CN254 - S0a, CN191 - Sa), and the two positive slope galaxies
have this morphology as well (CN849 \& CN228 - both Sab's). The other negative slope
galaxy, CN22, has a peculiar morphology and is clearly a merging system (which at
least is consistent with its negative H$\delta$ slope; see below). 
Two of the three flat-sloped galaxies, CN89 \& CN247, are both ellipticals. The only other
flat-sloped galaxy, CN143, has an Sbc morphology, which we note is the latest Hubble type in our 
E+A sample. Finally, it should also be noted that merger/interaction activity appears
to be prevalent amongst the positive and flat slope galaxies, with the $HST$-based imaging of 
\citet{couchw98} providing quite clear evidence that CN849 is interacting with its 
close, bright neighbour (with the two being connected by a tidal bridge), and somewhat
more tentative evidence that CN247 is involved in a similar interaction with its neighbour.
Although CN849 is an interacting galaxy (It is clearly not in the late stage of a merger) this
does not rule out that this galaxy has had its star formation recently truncated - producing
its E+A spectral signature.  

\subsection{Metallicity gradients}
Before we interpret these different radial H$\delta$ slopes further, it is important
that we first check whether some of this variation could be due to metallicity gradients
which are well known in galaxies. The strength of the hydrogen Balmer lines depends on
the main-sequence turnoff temperature. Young main-sequence stars are hot and 
have strong hydrogen absorption. Metal-poor stars also have relatively hot main-sequence 
stars and stronger hydrogen absorption. However the effect is small compared 
with the effect of age \citep{wortheyg97}. Metal absorption lines within the 
index continuum bands will also affect the derived value of the index.
Unfortunately, we cannot attempt a rigorous treatment of 
metallicity effects for these specific galaxies, because this would require further observation
with a broader spectral range and higher signal-to-noise ratios.  We can only discuss general trends.
The expectation here is that the metallicity  
decreases with increasing distance from the galactic centre \citep{searlel71}. 
There have been various suggestions for the  origin of the metallicity gradients 
in galaxies. The models of \citet{larsonr74,carlbergr84} predict that in the 
process of galaxy formation, the transfer and enrichment of gas toward the galactic centre leads to more 
metal enriched stars forming in the central regions and hence the 
development of a metallicity gradient. \citet{martinellia98} 
reproduced the observed metallicity gradients in galaxies using a model 
in which supernova-driven galactic winds develop first
in the outskirts of a galaxy and then successively progress toward the centre. 
\par
Since the equivalent width of the H$\delta$ line depends on 
metallicity as well as age, we need to evaluate the effect of a 
metallicity gradient on the H$\delta_{\mathrm{F}}$  profiles in Fig.~8. We cannot measure the 
metallicity gradients directly from our spectra because the observed wavelength 
range does not include a suitable metallicity index. Instead we used 
the models of \citet{thomasd04} to compute the effect of metallicity on our 
measured H$\delta_{\mathrm{F}}$  values. By assuming [$\alpha$/Fe]=0 and a single age population, a linear 
fit to these models yields the slope:
\begin{equation}
{ {d{\rm  H} \delta_{\mathrm{F}} } \over d{\rm [Z/H]}}\propto -1.98 {\rm \;\AA},
\end{equation}
which is essentially independent of the age of the stellar population. 
Typical metallicity gradients in early-type galaxies are:
\begin{equation}
{{d[Z/H]} \over d{\rm log} r } \propto -0.22\pm 0.09
\end{equation}
\citep{daviesr93}.  
We can combine equations (9) and (10) to estimate the radial variation in H$\delta_{\mathrm{F}}$  
which would result solely from a gradient in metallicity: 
\begin{equation}
{H\delta_{\mathrm{F}} } \propto (0.19{\rm \AA}) ln r+{\rm constant}
\end{equation}
Therefore, typical metallicity gradients in galaxies 
(in which the metallicity decreases with radius) result in an H$\delta_{\mathrm{F}}$  
signature that increases with radius. The radial profiles in Fig.~8 span a 
range in galacto-centric radii of approximately 1 to 5 kpc; from 
Equation (11), a fractional increase of 31\% in H$\delta_{\mathrm{F}}$  over 
this range would be expected from the metallicity gradient in Equation (10), or an average  
linear slope of $8{\rm \% \;  kpc^{-1}}$. In general this is small compared with 
the size of the observational errors for the E+A's with flat or  positive slopes (see Table 2). Of course for 
those E+A galaxies with negative slopes, a metallicity gradient only serves to 
diminish this type of radial behaviour; correcting for it would therefore just further 
steepen the increase in H$\delta$ absorption towards their centres.

\subsection{Comparison with numerical models}
In order to derive physical meaning from the observed radial gradients of
H$\delta_{\mathrm{F}}$  in E+As, we perform a set of numerical simulations of E+A galaxy formation.
We investigate two possible scenarios of E+A galaxy formation: (1)
Galaxy merging producing strong starbursts \citep[e.g., \citetalias{zabludoffa96};][]{bekkik98,bekkik01} and
(2) abrupt truncation of star formation in disk galaxies 
\citep[e.g.,][]{poggiantib96,shioyay04}. 

\subsubsection{Numerical simulations of  H$\delta$ gradients}
Since the numerical methods and techniques we employ for modelling the chemodynamical
and photometric evolution of galaxy mergers and truncated spirals
have already been described in detail 
elsewhere \citep{bekkik98b,bekkik01,bekkik02}, we give only  a brief review here. 

The progenitor disk galaxies that take part in a merger are taken to 
have a dark halo, a bulge, and a thin exponential disk
with a total disk mass ($M_{\rm d}$) of   6.0 $\times$ $10^{10}$ $ \rm M_{\odot}$ and
size ($R_{\rm d}$) = 17.5\,kpc. 
We adopt the density distribution of the NFW 
halo \citep{navarroj96} suggested from CDM simulations and 
the radial ($R$) and vertical ($Z$) density profiles 
of the  disk are  assumed to be
proportional to $\exp (-R/R_{0}) $ with scale length $R_{0}$ = 0.2$R_{\rm d}$,
and to  ${\rm sech}^2 (Z/Z_{0})$ with scale length $Z_{0}$ = 0.04$R_{\rm d}$,
respectively.
The disk is composed both of gas and stars with a gas mass fraction
($f_{\rm g}$) of 0.1. The gas is 
represented by a collection of discrete gas clouds that follow the observed mass-size
relationship. 
Field star formation
is modelled by converting  the collisional gas particles
into  collision-less new stellar particles. We adopt the Schmidt law \citep{schmidtm59}
with exponent $\gamma$ = 1.5 \citep[1.0  $ < $  $\gamma$ $ < $ 2.0,][]{kennicuttr98}
as the controlling
parameter of the rate of star formation. 
We determined the radial gradient of H$\delta$ equivalent width for a 
remnant of a merger scenario (with a prograde-retrograde orbital configuration)
at the post-starburst phase (more than 0.2\,Gyr after the peak of the starburst).

For the truncation disk model, to be self-consistent we use the same disk parameters as outlined above.
The bulge to disk ratio is 0.5. 
Star formation is {\it simultaneously and uniformly}
truncated throughout the entire disk after $\sim$ 2\,Gyr of isolated evolution (i.e., no
tidal interaction/merging and no triggered starbursts). 
The resulting disk can contain  
a large fraction of young stars
(i.e., A-type stars). 
Although our assumption of uniform truncation may be oversimplified, particularly 
in the case of truncation resulting from cluster-related 
processes such as ram-pressure stripping of disk gas and halo gas stripping, it is 
adequate for the present study, and serves to illustrate the 
remarkable differences in the time evolution of H$\delta$ equivalent width radial gradients between 
the two E+A models (i.e., merger-induced starburst  vs truncation).

\subsubsection{Comparison with observed H$\delta$ gradients}
The simulated H$\delta$ radial profiles for the galaxy merger and instantaneous truncation models are shown in Fig.~9 and Fig.~10
respectively. For the galaxy merger model in Fig.~9 we over-plot the H$\delta_{\mathrm{F}}$  measurements for our three negative-gradient galaxies:
CN254, CN22 and CN191 (top panel, centre panel and lower panel, respectively).  For the instantaneous truncation model we over-plot the observed 
H$\delta_{\mathrm{F}}$  values of three of our galaxies with positive and flat gradients: CN849, CN247 and CN228.  
In both cases we have displayed the  H$\delta_{\mathrm{F}}$ values corrected for convolution with the seeing disk as described in Section 4.3.  The
errors in the equivalent width measurements after deconvolution were approximated by
separately deconvolving the $\pm 1\sigma$ fits to the convolved radial equivalent width profile and comparing the results with the best fit.
The observed H$\delta_{\mathrm{F}}$  measurements are plotted 
as {\it filled squares} and the numerical model profiles as {\it solid lines}.  For the merger model (Fig.~9) we show the radial H$\delta$
 gradient of the merger remnant at times of 0.2, 0.75 and 1.5\,Gyrs since the maximal starburst. In the case of the instantaneous truncation model displayed
in Fig.~10 we show the H$\delta$ profiles at times 0.2, 0.5 and 1.5\,Gyrs after the truncation of star formation.
\par
It is interesting to compare the differences in the radial behaviour of the H$\delta$ absorption line between the merger and truncation models.
In the merger model a centralised burst of star formation is produced, when the starburst ends the galaxy is left with a central population of 
young stars and hence a radial distribution of H$\delta$ equivalent width which is highest in the centre and decreases rapidly with galacto-centric
radius (see $T_{PSB}=0.2$\,Gyr profile in Fig.~9).  As time progresses after the cessation of star formation the contribution to the integrated 
light from the youngest stars is diminished, resulting in a decrease in the H$\delta$ equivalent width which causes the radial profile to flatten.  At a time
 $T_{PSB} \approx 1.5$\,Gyr  after the maximal starburst the radial H$\delta$ profile has evolved to be flat and uniformly low across the 
entire extent of the galaxy -- the E+A galaxy signature is no longer present.  The H$\delta$ profiles shown in Fig.~9 are for a major merger 
model with a specific orbital configuration. However, it is expected that all tidally induced starbursts will exhibit qualitatively similar behaviour,
with the star formation concentrated in the galactic centre, producing a post-starburst galaxy spectrum with a radial H$\delta$ equivalent width profile which 
decreases with galacto-centric radius.  In contrast, the radial behaviour of the H$\delta$ absorption line
in the instantaneous truncation model is quite different. Immediately after the truncation of star formation the galaxy has a flat 
uniformly high H$\delta$ equivalent width profile (see $T_{Tr}=0.2$\,Gyr H$\delta$ equivalent width profile in Fig.~10).  The galaxy CN143 exhibits 
such a profile (compare the radial profile for CN143 in Fig.~8 and $T_{Tr}=0.2$\,Gyr profile in Fig.~10).  As the system evolves and the age of the 
youngest stellar population increases the contribution of the light from this young population of stars decreases. The decrease is most prominent 
in the centre of the galaxy where there exists a larger fraction of old stars; this results in the H$\delta$ equivalent width decreasing most rapidly in the 
central regions of the galaxy, leading to a positive H$\delta$ equivalent width gradient which steepens with time.
\par
Given the very different expectations for the behaviour and time evolution of the  H$\delta$ equivalent width radial profile in E+A galaxies for the
two formation mechanisms discussed (i.e. merger induced starbursts and abrupt truncation of star formation) we propose that the 
radial distribution of H$\delta$ absorption in E+A's provides a useful tool in differentiating between these two formation mechanisms.  In 
the sample of eight E+A galaxies examined here we find three galaxies (CN254, CN22 and CN191) which have a negative H$\delta_{\mathrm{F}}$  gradient.
We suggest that this radial behaviour provides evidence that the E+A spectral signature in these galaxies has 
its origin in a merger or tidal interaction. The remaining five E+A galaxies (CN143, CN89, CN849, CN228 and CN247) have either flat or radially increasing 
H$\delta_{\mathrm{F}}$  profiles, these profiles are more consistent with the recent truncation of star formation in normal disk galaxies.

\section{Summary}
We have obtained IFU spectra for 12 galaxies in AC114 originally classified as E+A systems by \citetalias{couchw87}, permitting the
first investigation of the spatial distribution of H$\delta$ absorption in such galaxies. We summarise our findings as follows:

\begin{list}{$\bullet$}{}
\item We find global H$\delta$ equivalent width values lower than in the original \citetalias{couchw87} sample from which our 
targets were selected. This is especially the case for the five galaxies which were original assigned an HDS classification by \citetalias{couchw87}.
We attribute this to differences in aperture placement and/or systematic errors in the sky subtraction of the original \citetalias{couchw87} sample.
Some galaxies appear to have been misclassified based on the original \citetalias{couchw87} spectroscopy.

\item Three galaxies in our sample exhibit negative H$\delta_{\mathrm{F}}$  gradients implying a strong central concentration of young stars. The E+A 
galaxies in our sample which exhibit this property are CN254, CN22 and CN191. Normal metallicity gradients in galaxies, in which metallicity 
decreases with galactic radius, would imply an even more significant central concentration, as does correction for the smearing of the galaxy 
light due to seeing. The distribution of young star light in these galaxies is consistent with the origin of the global E+A signature being the 
result of a merger or tidal interaction.

\item Three galaxies in the sample, CN143, CN89 \& CN247 (which has globally low H$\delta_{\mathrm{F}}$ ) have a H$\delta_{\mathrm{F}}$ profile
which is statistically consistent with being flat. This profile is consistent with a galaxy observed shortly after the 
global truncation of its star formation.  

\item Two galaxies in the sample show a positive H$\delta_{\mathrm{F}}$  gradient with H$\delta$ absorption being at its weakest at the centre of the galaxy.
The galaxies in this category are CN89 \& CN228. These H$\delta_{\mathrm{F}}$  
profiles are consistent with a galaxy which has undergone a global truncation of its star formation within the last 2\,Gyrs.
\end{list}

\def\ref{\par\noindent\hangindent\parindent}

\section*{Acknowledgements}
M.B.P. was supported by an Australian Postgraduate Award.
W.J.C., C.B., and K.B. acknowledge the financial support of the Australian
Research Council throughout the course of this work.  
We wish to thank the referee, Tomotsugu Goto, for a helpful and thorough 
report which has greatly improved this paper.  This research has made use of the NASA/IPAC 
Extragalactic Database (NED) which is operated by the Jet Propulsion Laboratory, California 
Institute of Technology, under contract with the National Aeronautics and Space Administration. 

\bibliographystyle{mn2e}
\bibliography{references}

\setcounter{figure}{8}
\begin{figure*}
{\includegraphics[width=18cm,angle=0]{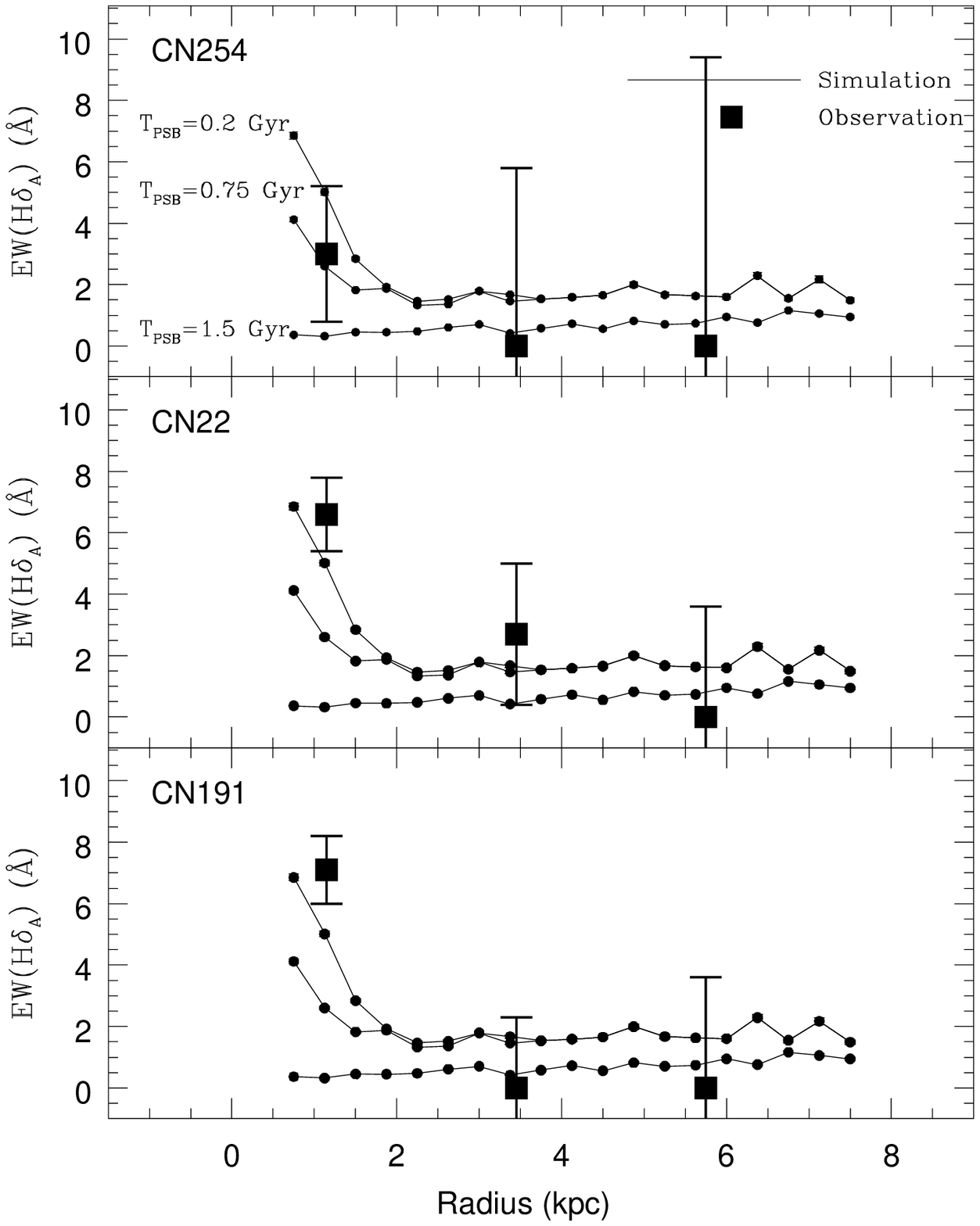}}
\caption{Model H$\delta$ radial profiles for our merger model ({\it solid lines}).  The model curves are shown with 
with the observed (deconvolved) H$\delta_{\mathrm{F}}$  data for CN254 ({\it top panel}), CN22 ({\it middle}) and 
CN191 ({\it lower}) superimposed as {\it filled squares}.}
\end{figure*}

\setcounter{figure}{9}
\begin{figure*}
{\includegraphics[width=18cm,angle=0]{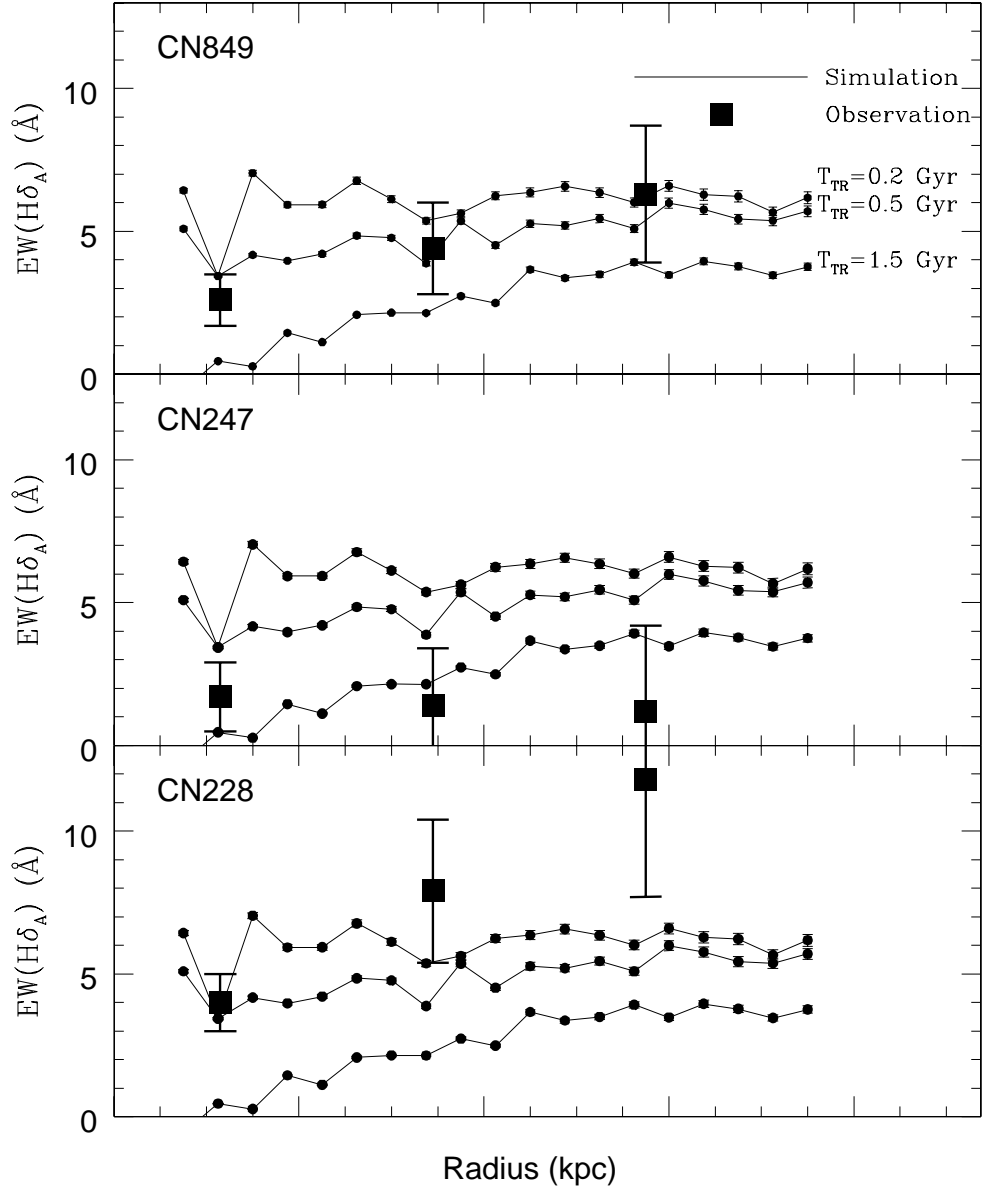}}
\caption{Model H$\delta$ radial profiles for our instantaneous truncation model ({\it solid lines}).  The model curves are shown with 
with the observed (deconvolved) H$\delta_{\mathrm{F}}$  data for CN849 ({\it top panel}), CN247 ({\it middle}) and 
CN228 ({\it lower}) superimposed as {\it filled squares}.}
\end{figure*}

\label{lastpage}

\end{document}